\documentclass[pra,twocolumn,showpacs,preprintnumbers,amsmath,amssymb]{revtex4}
\usepackage{color}
\usepackage{graphicx}
\usepackage{dcolumn}
\usepackage{bm}

\newcommand{\vek}[1]{\bm{\mathrm{#1}}}
\newcommand{\ave}[1]{\langle #1\rangle}
\newcommand{\av}{\vek{a}}
\newcommand{\kv}{\vek{k}}
\newcommand{\pv}{\vek{p}}

\newcommand{\qv}{\vek{q}}
\newcommand{\rv}{\vek{r}}

\newcommand{\vv}{\vek{v}}
\newcommand{\tautilde}{\tilde{\tau}}
\newcommand{\Ntil}{\tilde{N}}
\newcommand{\calO}{\mathcal{O}}

\newcommand{\nablav}{\bm{\nabla}}
\newcommand{\eq}{\mathit{eq}}
\newcommand{\pot}{\mathit{pot}}
\newcommand{\kin}{\mathit{kin}}

\newcommand{\ho}{\mathit{ho}}
\newcommand{\coll}{\mathit{coll}}

\newcommand{\Eq}[1]{Eq.\@ (\ref{#1})}
\newcommand{\Eqs}[1]{Eqs.\@ (\ref{#1})}
\newcommand{\Ref}[1]{Ref.\@ \cite{#1}}
\newcommand{\Refs}[1]{Refs.\@ \cite{#1}}
\newcommand{\Fig}[1]{Fig.\@ \ref{#1}}

\newcommand{\Sec}[1]{Sec.\@ \ref{#1}}

\DeclareMathOperator{\Imag}{Im}

\renewcommand{\Im}{\Imag}
\begin{document}


\title{Numerical solution of the Boltzmann equation for the collective
modes of trapped Fermi gases.}

\author{Thomas Lepers}
\email{t.lepers@ipnl.in2p3.fr}
\affiliation{Universit{\'e} de Lyon, F-69622 Lyon, France;
  Univ. Lyon 1, Villeurbanne;
  CNRS/IN2P3, UMR5822, IPNL}
\author{Dany Davesne}
\affiliation{Universit{\'e} de Lyon, F-69622 Lyon, France;
  Univ. Lyon 1, Villeurbanne;
  CNRS/IN2P3, UMR5822, IPNL}
\author{Silvia Chiacchiera}
\affiliation{Centro de F{\'i}sica Computacional, Department of Physics,
University of Coimbra, P-3004-516 Coimbra, Portugal}
\author{Michael Urban}
\affiliation{Institut de Physique Nucl{\'e}aire, CNRS/IN2P3 and
  Universit\'e Paris-Sud 11, 91406 Orsay Cedex, France}
\date{April 29, 2010}
\begin{abstract}
We numerically solve the Boltzmann equation for trapped fermions in
the normal phase using the test-particle method. After discussing a
couple of tests in order to estimate the reliability of the method, we
apply it to the description of collective modes in a spherical
harmonic trap. The numerical results are compared with those obtained
previously by taking moments of the Boltzmann equation. We find that
the general shape of the response function is very similar in both
methods, but the relaxation time obtained from the simulation is
significantly longer than that predicted by the method of moments. It
is shown that the result of the method of moments can be corrected by
including fourth-order moments in addition to the usual second-order
ones and that this method agrees very well with our numerical simulations.
\end{abstract}
\maketitle

\section{\label{sec:intro}Introduction}
In experiments on ultracold trapped Fermi gases, there are many
situations where the system is out of thermal equilibrium. The first
one is of course the trapping and cooling stage, i.e., before the
system has reached its equilibrium state which is usually the starting
point for the actual experiment. Then, in some experiments the system
is excited in order to observe its dynamical behavior. For instance,
many experiments studied collective oscillations of the system
\cite{KinastHemmer,KinastTurlapov,Bartenstein,Altmeyer_quadrupole,Altmeyer_breathing,RiedlBruun,Nascimbene},
another example being a recent experiment at MIT where the collision of
two atom clouds (both in equilibrium) was studied
\cite{Zwierlein}. Finally, often the system is not imaged directly
during the experiment, but only after the trap was switched off and
the system has expanded for a certain time, in order to increase its
size.

The modeling of such time-dependent processes from the theoretical
point of view can be quite complicated. For practical reasons, only
semiclassical approaches are suitable for the description of
time-dependent phenomena involving typically several $10^5$ atoms in a
three-dimensional, non-uniform geometry. In some cases, it is possible
to use hydrodynamic approaches: Superfluid hydrodynamics describes the
expansion \cite{Menotti} and the collective modes \cite{Cozzini} of
superfluid systems at zero temperature. Hydrodynamics is also
applicable in the normal-fluid phase if the mean time between
collisions is much shorter than all other time scales of the process
under consideration, so that the system can always be considered to be
in a local equilibrium \cite{Cozzini,Pedri}. Superfluid and normal
hydrodynamics can be combined to two-fluid hydrodynamics in order to
describe superfluid systems at finite temperature
\cite{TaylorGriffin}. However, in many cases hydrodynamic approaches
are not sufficient. In all cases where it is important that even
locally the distribution $f(\rv,\pv,t)$ of the atoms is not an
equilibrium one, the Boltzmann equation allows a very general
description, provided the system is in the normal phase. If the system
is superfluid, a more elaborate theory is necessary which couples the
dynamics of the quasiparticle distribution function to the dynamics of the
superfluid order parameter \cite{UrbanSchuck,Urban}.

In the past, several authors used the Boltzmann equation for the
investigation of collective oscillations in normal-fluid trapped Fermi
gases
\cite{RiedlBruun,Pedri,ToschiVignolo,ToschiCapuzzi,MassignanBruun,BruunSmith,Chiacchiera}.
In most cases, the Boltzmann equation was not solved directly in order
to find the distribution function $f$, but semi-analytical approximate
solutions were found by using the scaling ansatz \cite{Pedri} or the
method of moments
\cite{MassignanBruun,BruunSmith,RiedlBruun,Chiacchiera}. These methods
rely explicitly or implicitly on the assumption that the collision
term can be treated in the relaxation-time approximation, with a
single relaxation time $\tau$ which is independent of the position in
the trap. An exception is the work by Toschi et
al. \cite{ToschiVignolo,ToschiCapuzzi}, where the Boltzmann equation
was solved numerically, using a test-particle method very similar to
the one we are using here. The test-particle method for the solution
of the Boltzmann equation has been used for many years in nuclear
physics for the simulation of heavy-ion collisions
\cite{BertschDasGupta}. In the context of trapped atoms, it was also
used for the simulation of the dynamics of the thermal cloud in a
Bose-Einstein condensate \cite{JacksonZaremba} and (without collision
term) of the normal-fluid component in a superfluid Fermi gas
\cite{Urban}.

For the collision term of the Boltzmann equation, it is important to know 
the cross section which in principle can be modfied by in-medium effects.
In the work by Riedl et al. \cite{RiedlBruun}, it was shown that
by using in the Boltzmann equation instead of the free cross-section
the in-medium one, the agreement between theoretical and experimental
frequencies and damping rates of different collective modes is
deteriorated. The reason is that the in-medium cross section is larger
than the free one, so that the relaxation time is dramatically
reduced. In our previous work \cite{Chiacchiera}, our aim was to
include in addition to the in-medium cross section medium effects into
the mean-field potential. This mean field resulted in better density
profiles and allowed us to understand the shift of the quadrupole
frequency in the collisionless regime at very low temperature observed
in \cite{Altmeyer_quadrupole}. However, it did not help to improve the
agreement between theory and experiment in the region of higher
temperatures, where the properties of the collective modes are
completely dominated by collisions.

Hence, one of our motivations for the present work was to check the
validity of the relaxation-time approximation which is implicitly made
in the method of moments. Here we will restrict ourselves to the case
without mean field and with the free cross section. As we will show in
\Sec{sec:Modes}, the numerical solution of the Boltzmann equation
gives indeed a significantly longer relaxation time than the method of
moments. As we will show, this discrepancy is due to the
restriction of the method of moments to second-order moments in the
existing literature
\cite{MassignanBruun,BruunSmith,RiedlBruun,Chiacchiera}. Once
fourth-order moments are included, the results of the method of
moments and of the numerical solution are in good agreement.
However, already in the simplest case of a spherical harmonic trap
without mean field, the inclusion of fourth-order moments is a very
tedious task, while the numerical method can be generalized to
more realistic cases.

In addition, there are some other reasons why we felt the necessity for
a numerical method. For example, there are damping effects due to the
anharmonicity of the trap potential which cannot be described by the
method of moments. Another advantage of the numerical method is that
it offers the possibility to simulate not only the oscillation of the
cloud, but also the subsequent expansion after the trap has been
switched off.

In \Sec{sec:Numericaldescription} of the present paper, we give a
detailed description of the method. In particular, we explain in
detail how the collisions are simulated, since our method is somewhat
different from that of \Ref{ToschiVignolo}. Moreover we
discuss some tests we made in order to estimate up to which
precision we can trust our simulation. Then, in \Sec{sec:Modes}, we
come to the main point of our article and calculate the properties of
some collective modes for a system in a spherical harmonic trap. While
the sloshing and breathing modes are rather trivial, the
frequency and damping rate of the quadrupole mode are very sensitive
to the collisions. We compare the numerical results with those of the
method of moments. Finally, in \Sec{sec:Conclusions} we summarize and
give an outlook to future studies.

Throughout the paper, we will use units with $\hbar = k_B = 1$ ($\hbar
= $ reduced Planck constant, $k_B$ = Boltzmann constant). The strength
of the interaction is characterized by the dimensionless quantity $k_F
a$, where $a$ is the scattering length. Concerning the Fermi momentum
$k_F$ and the Fermi energy $E_F$ we follow the usual convention that
these quantities are defined by the corresponding ones of an ideal
Fermi gas at zero temperature, i.e., $k_F = \sqrt{2mE_F}$, $m$ being
the atomic mass, and $E_F = (3N)^{1/3}\omega_0$, where $N$ is the
number of atoms and $\omega_0$ the trap frequency. Temperatures will
be measured in units of the Fermi temperature $T_F = E_F$ (since $k_B
= 1$).
\section{Description of the numerical method}
\label{sec:Numericaldescription}
\subsection{Test-particle method}
\label{sec:Testparticlemethod}
We study a two-component ($\sigma = \uparrow,\downarrow$) gas of
fermionic atoms of mass $m$ in a potential $V(\rv,t)$ with attractive
interaction $a < 0$. 
We assume that the system is in the normal
phase and that it can be described semiclassically by phase-space
distribution functions $f_\sigma(\rv,\pv,t)$. In this paper, we will
restrict ourselves to the case that the distribution functions of both
spin states are equal ($f_{\uparrow}= f_{\downarrow}=f$), but the
generalization of the method to the cases of different distribution
functions, more than two components, or components with different
masses is straight-forward.

The time evolution of the distribution function $f$ is governed by the
Boltzmann equation \cite{Landau10}
\begin{equation}
\label{eq:eqboltzmann}
\dot{f}+ \dot{\rv}\cdot\nablav_r f+\dot{\pv}\cdot\nablav_p f=-I[f]\,,
\end{equation}
where the left-hand side (lhs) describes the particle propagation, with
\begin{equation}
\label{eq:eqmotion}
\dot{\rv} = \frac{\pv}{m}\quad \mbox{and}\quad \dot{\pv} = -\nablav V\,,
\end{equation}
and $I[f]$ on the right-hand side (rhs) denotes the collision term
which will be discussed later. 
The potential felt by the particles is the trap potential
that contains a static part and a time dependent one (which will be used 
to simulate the excitation of the collective modes) 
$V(\rv,t)=V_T(\rv)+V_1(\rv,t)$.

The density per spin state is related
to the distribution function by
\begin{equation}
\rho(\rv,t) = \int \frac{d^3p}{(2\pi)^3} f(\rv,\pv,t)\,,
\label{eq:density}
\end{equation}
and the number of atoms is given by
\begin{equation}
\label{eq:normalization}
N = N_{\uparrow} + N_{\downarrow} = 2 \int d^3r \rho(\rv,t)\,.
\end{equation}

The basic idea of the test-particle method (also called pseudoparticle
method) for solving the Boltzmann equation consists in replacing the
continuous distribution function by a sum of delta functions,
\begin{equation}
\label{eq:defdistri}
f(\textbf{r},\textbf{p},t)= \frac{N}{2\Ntil} \sum_{i=1}^{\Ntil}
  (2\pi)^3 \delta(\pv-\pv_i(t))\delta(\rv-\rv_i(t)) \,,
\end{equation}
where $\Ntil$ is the number of ``test particles''. This allows one to
express the average of an arbitrary single-particle observable
$F(\rv,\pv)$ in the simple form
\begin{equation}
\ave{F} = \frac{2}{N}\int \frac{d^3 r d^3p}{(2\pi)^3}
  f(\rv,\pv,t) F(\rv,\pv)
  = \frac{1}{\Ntil}\sum_{i=1}^{\Ntil} F(\rv_i,\pv_i)\,.
\end{equation}
In order to sample the six-dimensional phase space, it is necessary to
choose a sufficiently large number of test particles $\Ntil$ (usually
$\Ntil > N$). Neglecting the collision term $I[f]$ for the moment, it
is easy to see that \Eq{eq:defdistri} satisfies the Boltzmann equation
(\Eq{eq:eqboltzmann}) if the positions $\rv_i$ and momenta $\pv_i$ of
each test particle $i$ follow the classical equations of motion,
\Eq{eq:eqmotion}.

In practice, the delta functions in \Eq{eq:defdistri} can pose some
problems. For instance, they do not result in a continuous density
$\rho(\rv)$. Therefore it is often useful to replace them by Gaussians
of width $w_r$ and $w_p$ in position and momentum space, respectively:
\begin{equation}
\delta(\pv-\pv_i)\delta(\rv-\rv_i)\to 
  g_{w_p}(\pv-\pv_i)g_{w_r}(\rv-\rv_i)\,,
\label{eq:gaussians}
\end{equation}
with
\begin{equation}
g_{w_p}(\pv) = \frac{e^{-p^2/w_p^2}}{(\sqrt{\pi}w_p)^3}
\quad\mbox{and}\quad
g_{w_r}(\rv) = \frac{e^{-r^2/w_r^2}}{(\sqrt{\pi}w_r)^3}
\,.
\end{equation}
The widths $w_r$ and $w_p$ must be adapted such that they smooth out
the fluctuations due to the finite number of test particles, but not
the structure of the distribution function $f$. The statistical
fluctuations are of the order of $(2\Ntil w_r^3w_p^3/N)^{-1/2}$, i.e.,
the first condition is equivalent to
\begin{equation}
w_rw_p\gg\left(\frac{N}{2\Ntil}\right)^{1/3}\,.
\label{conditionN}
\end{equation}
The second condition implies of course that $w_r \ll R_{TF}$ and $w_p\ll
p_F$, where $R_{TF}$ and $p_F$ are the Thomas-Fermi radius and the
Fermi momentum, respectively, but this is not always
sufficient. At low-temperature, it is crucial to resolve the rapid
change of the distribution function around the Fermi surface, i.e.,
\begin{equation}
w_p\ll p_F \frac{T}{T_F}\quad\mbox{and}\quad w_r\ll R_{TF} \frac{T}{T_F}\,.
\label{conditionT}
\end{equation}
In practice, as the computation time increases as $\Ntil^2$, it turns
out that the conditions (\ref{conditionN}) and (\ref{conditionT})
cannot simultaneously be satisfied at too low temperatures.

\subsection{Particle propagation}
\label{sec:Particlepropagation}
In the absence of collisions, the numerical task consists only in
solving simultaneously the classical equations of motion
(\ref{eq:eqmotion}) for the $\Ntil$ test particles. We do this by
using the velocity Verlet algorithm \cite{Swope}, which contrary to
the original Verlet algorithm \cite{Verlet} uses the positions
$\rv_i(t_n)$ and velocities $\vv_i(t_n) = \pv_i(t_n)/m$ as starting
point for the time step from $t_n$ to $t_{n+1} = t_n+\Delta t$.
The propagation from $t_n$ to $t_{n+1}$ is done according to
\begin{gather}
\vv_i(t_{n+1/2}) = \vv_i(t_n) + \av_i(t_n)\Delta t/2\,\\
\rv_i(t_{n+1}) = \rv_i(t_n) + \vv_i(t_{n+1/2})\Delta t\,\\
\vv_i(t_{n+1}) = \vv_i(t_{n+1/2}) + \av_i(t_{n+1})\Delta t/2\,,
\end{gather}
where $\av_i(t) = -\nablav V(\rv_i(t),t)/m$ is the acceleration of the
$i$-th test particle. If it is written in this way, it is obvious that
the velocity Verlet algorithm is identical to the leap-frog algorithm
\cite{numericalrecipes}. Note that the accelerations $\av_i(t_{n+1})$
can be reused in the next time step, so that the algorithm needs only
one evaluation of the acceleration per time step, but nevertheless its
global error is of the order $\calO(\Delta t)^2$. This allows us to
obtain a good accuracy for reasonable time steps $\Delta t$. A good
test of the particle propagation is to check the energy conservation:
typically we find $|E_i(t)-E_i(0)|/E_i(0) \simeq 10^{-6}$ for each
test particle and for all times considered.
\subsection{Collision term}
\label{sec:Collisionterm}
The rhs of the Boltzmann equation (\ref{eq:eqboltzmann}) describes the
collisions between particles of opposite spin. It thus depends on the
scattering cross section $d\sigma/d\Omega$ and reads \cite{Landau10}
\begin{multline}
I[f]=\int \frac{d^3 p_1}{(2 \pi)^3} \int d\Omega \frac{d\sigma}{d\Omega} 
|\vv-\vv_1| [f f_1 (1-f^\prime)(1-f_1^\prime)\\
  -f^\prime f_1^\prime (1-f) (1-f_1)] \,.
\label{eq:collisionterm}
\end{multline}
In the first term, $\pv$ and $\pv_1$ are the incoming momenta,
$\pv^\prime$ and $\pv_1^{\prime}$ are the outgoing ones, $\Omega$ is
the solid angle formed by the incoming relative momentum $\pv-\pv_1$
and the outgoing relative momentum $\pv^{\prime}-\pv_1^{\prime}$, and
$f \equiv f(\rv,\pv,t)$, $f_1 \equiv f(\rv,\pv_1,t)$, etc. In the
second term, the role of incoming and outgoing momenta is
exchanged. Momentum and energy conservation implies $\pv + \pv_1 =
\pv^\prime + \pv_1^\prime$ and $|\pv-\pv_1| =
|\pv^\prime-\pv_1^\prime|$. Here we consider the case of pure $s$-wave
scattering, in which the cross section is isotropic, i.e.,
$d\sigma/d\Omega = \sigma/4\pi$. In principle the cross section is
modified by medium effects \cite{BruunSmith,Chiacchiera}, but in the
present paper we will only use the free cross-section (i.e., the
cross-section for the scattering of two atoms of opposite spin in free
space) which is given by \cite{Landau3}
\begin{equation}
\sigma = \frac{4\pi a^2}{1+(qa)^2}\,,
\end{equation}
where $q = |\pv-\pv_1|/2 = |\pv^\prime-\pv_1^\prime|/2$.

In our numerical simulation, the collision term is included by
allowing the test particles to collide with each other. The cross
section of the test particles, $\tilde{\sigma}$, is related to the
cross section of the atoms by $\tilde{\sigma} = \sigma N/2\Ntil$
(since $\Ntil$ test particles represent $N/2$ atoms of a given
spin). Whether a pair $i,j$ of test particles collides in a time step
$t_n$ or not is determined as follows: First, we determine if the two
particles are at their closest approach in the present time
step. Explicitly, if we write $\rv_{ij} = \rv_i-\rv_j$ and $\vv_{ij} =
\vv_i-\vv_j$, the closest approach is reached at $t_\mathit{min} =
t_n-\rv_{ij}\cdot\vv_{ij}/v^2_{ij}$ and we check if
$|t_\mathit{min}-t_n| < \Delta t/2$. If yes, we calculate the
corresponding minimal distance by $d_\mathit{min}^2 =
r_{ij}^2-(\rv_{ij}\cdot\vv_{ij})^2/v_{ij}^2$ and check if $\pi
d_\mathit{min}^2 < \tilde{\sigma}$. In this case, the collision is
classically allowed. We then propagate both test particles to
$t_\mathit{min}$, change the direction of their relative velocity
$\vv_{ij}$ in a random way (thus conserving the total momentum and the
total energy), and propagate them back to the original time
$t_n$. Finally, in order to take into account the Pauli-blocking
factors in \Eq{eq:collisionterm}, we calculate the occupation numbers
$f_i^\prime$ and $f_j^\prime$ at the new positions and momenta
($f_i^\prime = f(\rv_i^\prime,\pv_i^\prime)$ etc.) using
\Eq{eq:defdistri} with Gaussians instead of delta functions, see
\Eq{eq:gaussians}. With probability $(1-f_i^\prime)(1-f_j^\prime)$
the collision is allowed and we keep the new positions and momenta,
otherwise the collision is blocked and we keep the old ones.

We checked that the total energy is still well conserved when
collisions are switched on: typically we find better than
$|\ave{E}(t)-\ave{E}(0)|/\ave{E}(0) \simeq 10^{-5}$ for all times
considered.
\subsection{Initialization}
\label{sec:Initialization}
Before the simulation can start, the test-particle positions and
momenta have to be initialized. Here we assume that the system is
initially in equilibrium.

A suitable equilibrium distribution is given by the distribution
function within the Thomas-Fermi or local-density approximation (LDA),
\begin{equation}
f_\eq(\rv,\pv) = \frac{1}{e^{\beta(p^2/2m+V_T(\rv)-\mu)}+1}\,,
\label{eq:equilibrium}
\end{equation}
since it is a stationary solution of the Boltzmann equation
\cite{Landau10}. This distribution has two parameters, namely the
inverse temperature $\beta = 1/T$ and the chemical potential
$\mu$. The temperature $T$ is an input parameter, whereas the chemical
potential $\mu$ is determined by demanding that the integral of
\Eq{eq:equilibrium} over $\rv$ and $\pv$ gives the right number of
atoms. 

Having determined the chemical potential $\mu$, we randomly generate
the test-particle positions and momenta in such a way that the
probability to be at position $\rv$ and to have momentum $\pv$ is
proportional to $f_\eq(\rv,\pv)$. In practice, we do this by
first generating the positions according to the density profile
obtained from \Eq{eq:density} with $f=f_\eq$. Then we generate
the momenta according to $f_\eq$.

\subsection{Tests of reliability and accuracy}
\label{sec:Tests}
In this subsection we describe two main tests we made to 
be sure that our code is reliable. 
Here, we assume the potential to be static and, 
as in the rest of the paper, 
we use a spherical harmonic potential 
\begin{equation}
V(\rv,t) = V_T(\rv) = \frac{1}{2} m\omega_0^2 r^2\,.
\label{eq:harmonic}
\end{equation}
This potential defines naturally a time scale $1/\omega_0$, a length
scale $l_\mathit{ho} = 1/\sqrt{m\omega_0}$, an energy scale
$\omega_0$, and so on.

Let us consider the energy distribution of the atoms,
\begin{equation}
\frac{dN}{dE} = 2\int\frac{d^3 r\,d^3 p}{(2\pi)^3}
  f(\rv,\pv)\delta\Big(\frac{p^2}{2m}+V_T(\rv)-E\Big)\,.
\label{eq:distrib}
\end{equation}
In equilibrium, the distribution should be given by $dN/dE =
g(E)/(e^{(E-\mu)/T}+1)$, where $g(E)$ is the density of states
(including the degeneracy factor 2). In the present case of a
spherical harmonic oscillator, we have $g(E) = E^2/\omega_0^3$. In the
absence of collisions, energy conservation automatically implies that
the distribution stays constant, but in the presence of collisions
this test is a non-trivial check of the Pauli blocking in the
simulation. Within the test particle method, $dN/dE$ is obtained by
counting the test particles in energy bins.

\begin{figure}
\begin{center}
\includegraphics[width=8cm]{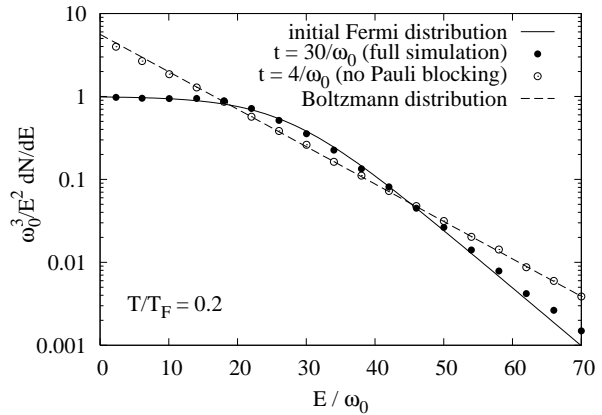}
\end{center}
\caption{\label{fig:distri} Energy distribution of the atoms 
(divided by the density of states) for $T/T_F = 0.2$ (see text
for details).
The system consists of $N = 10000$ atoms with a scattering length
$a = -0.2537 l_\ho$ ($1/(k_F a) = -0.5$). 
The parameters of the simulation are: $\Ntil = 50000$, 
$w_r = 1.5 l_{\ho}$, $w_p = 1.5/l_{\ho}$, and $\Delta
t = 0.02/\omega_0$.}
\end{figure}

In \Fig{fig:distri}
we show, for $T/T_F = 0.2$, the initial Fermi distribution (solid
line) and the stationary distribution obtained in the numerical
simulation after $t = 30/\omega_0$ (filled circles). The agreement
between the distribution generated by the simulation and the initial
Fermi one is not perfect, but satisfactory. In order to show that this
is not a trivial result, let us see what happens if we switch off the
Pauli blocking in the simulation of the collision term. In this case,
already after a relatively short time $\sim 3/\omega_0$, the
distribution in the numerical simulation (empty circles) has converged
to a Boltzmann distribution with the same number of atoms and total
energy (dashed line). So, the stability of the Fermi distribution in
our full simulation shows clearly that Pauli blocking is correctly
implemented. The small deviations from the ideal Fermi distribution
are a consequence of the fact that with the chosen widths of the
Gaussians ($w_r = 1.5 l_{\ho}$ and $w_p = 1.5/l_{\ho}$), the condition
(\ref{conditionT}) is not well satisfied at $T/T_F = 0.2$. When we did
the same kind of comparison at higher temperatures, we found that the
agreement between the simulation and the Fermi distribution improves:
at $T/T_F=0.4$, it is already perfect.

The test described above is independent of the actual number of
collisions. In order to check the latter, let us look at the collision
rate
\begin{multline}
\label{eq:collrate}
\dot{N}_\coll = \int d^3r \int \frac{d^3 p}{(2 \pi)^3} 
\int \frac{d^3 p_1}{(2 \pi)^3} \int d\Omega \frac{d\sigma}{d\Omega}
|\vv-\vv_1|\\ \times f f_1(1-f^{\prime})(1-f_1^{\prime}) \,.
\end{multline}
In the numerical simulation, this quantity can be obtained as
$\dot{N}_\coll = (N/\Ntil) \dot{\Ntil}_\coll$, where
$\dot{\Ntil}_\coll$ denotes the number of collisions of test particles
per unit time.

Although in equilibrium the net effect of collisions is zero, the
collision rate in equilibrium is a good test for the simulation
because it can be compared with the exact result 
[\Eq{eq:MCblock}, see Appendix \ref{app:collrate}].
For testing purposes, 
it is useful to compare also the total rate of allowed and blocked
collisions with the exact result [\Eq{eq:MCnoblock}]. 

\begin{figure}
\includegraphics[width=8cm]{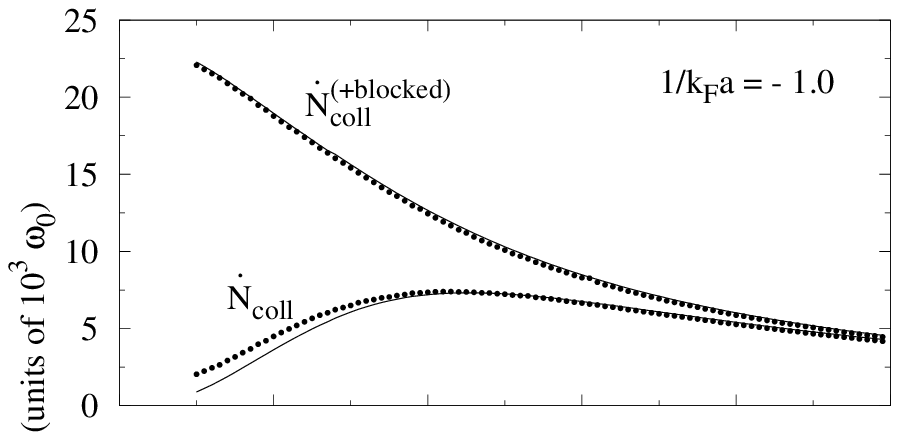}\\[-3mm]
\includegraphics[width=8cm]{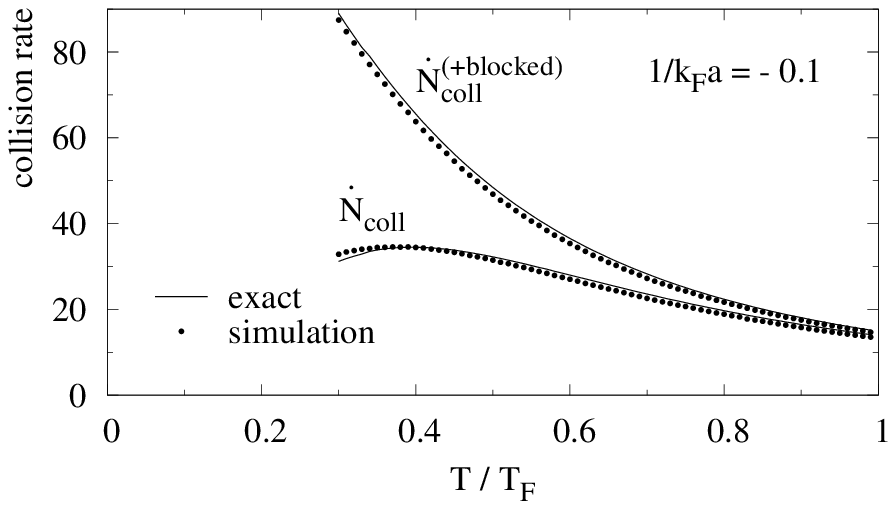}
\caption{Simulated collision rates (filled circles) with and without
  blocking compared with the corresponding exact results,
  \Eqs{eq:MCblock} and \Eq{eq:MCnoblock} (solid lines), for a gas of
  10000 atoms with interaction strength $1/k_F a = -1.0$ (top) and
  $-0.1$ (bottom).}
\label{fig:collrate}  
\end{figure}

In \Fig{fig:collrate}, the collision rates (with and without blocking)
of the simulation are shown together with the exact results as
functions of the temperature for two different values of the
scattering length. In the case of relatively weak interaction, $1/k_Fa
= -1$ (upper panel), we see that the agreement between the simulation
and the exact result is excellent for temperatures above $\sim 0.35 \,
T_F$. Below that temperature, the collision rate in the simulation
with Pauli blocking gradually becomes too high since the finite widths
of the Gaussians ($w_r = 1.5l_\ho$, $w_p = 1.5/l_\ho$) do not satisfy
any more the condition (\ref{conditionT}) and act in the
Pauli-blocking factors like an enhanced temperature. In the rest of
this paper, we will therefore restrict ourselves to temperatures above
$0.2 \, T_F$. Near unitarity ($1/k_F a = -0.1$), we consider only
temperatures above $0.3 \, T_F$ because this is close to the
superfluid transition temperature at unitarity \cite{Chiacchiera}. As
it can be seen in the lower panel of \Fig{fig:collrate}, the agreement
between the collision rate obtained in the simulation and the exact
one is satisfactory in the temperature range considered. The agreement
is not as good as for $1/k_Fa = -1$ at high temperature because of the
larger cross section which leads to collisions between test particles
which are further apart.

\section{Simulation of collective modes}
\label{sec:Modes}

\subsection{Sloshing mode}
\begin{figure}
\includegraphics[width=8cm]{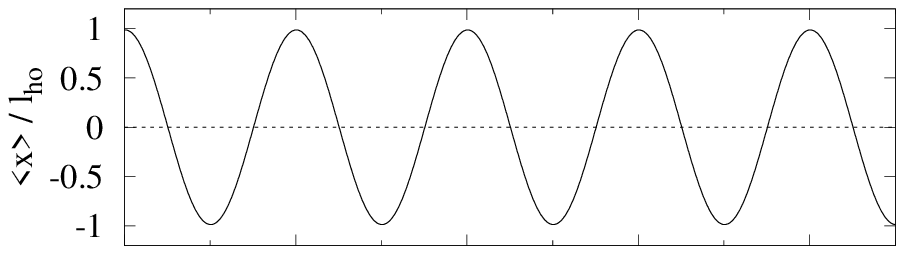}\\[-4mm]
\includegraphics[width=8cm]{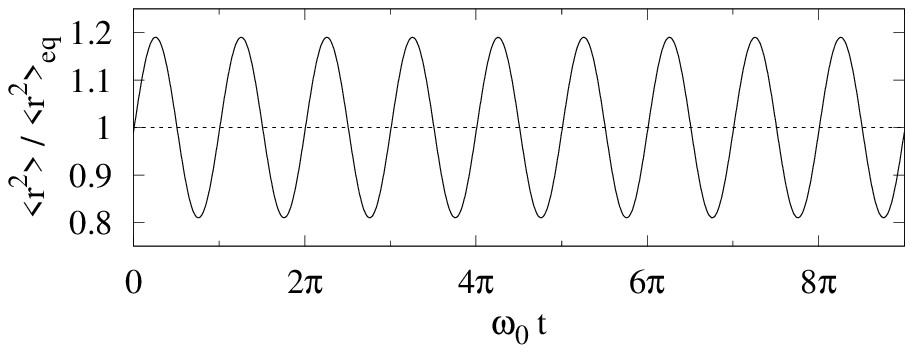}
\caption{Top: simulation of the sloshing mode. The mode was excited at
  $t = 0$ by displacing all test particles by $l_{ho}$ in the $x$
  direction. Bottom: simulation of the breathing mode. The mode was
  excited by changing at $t = 0$ all test-particle momenta according
  to $\pv_i\to\pv_i+c\rv_i$ ($c=0.2 m \omega_0$). Both
  simulations were done for a system of $N=5000$ particles at $T=0.4 \,
  T_F$ and $1/k_Fa = -0.3$.}
\label{fig:sloshingbreathing}  
\end{figure}

The sloshing mode is an oscillation of the center of mass of the
system. It plays a special role because in a harmonic trap it is
undamped and its frequency is equal to that of the trap, independently
of the number of atoms, of the temperature, and of the interaction
between the atoms (Kohn mode \cite{Kohn,Brey}). This is why it is
often used for the experimental determination of the trap frequency
\cite{Altmeyer_breathing}. Within the test-particle method, this
general theorem is satisfied and it is easy to see why:

Let us first neglect collisions. From the equations of motion of the
individual test particles in the harmonic potential
(\ref{eq:harmonic}),
\begin{equation}
\label{eq:eqmotionharmonic}
\dot{\rv}_i = \pv_i/m \quad \mbox{and}\quad \dot{\pv}_i =
-m\omega_0^2\rv_i
\end{equation}
it is evident that the averages $\ave{\rv}$ and $\ave{\pv}$ obey
analogous equations of motion,
\begin{equation}
\label{eq:eqmotionsloshing}
\frac{d}{dt}\ave{\rv} = \frac{\ave{\pv}}{m}\quad \mbox{and}\quad
\frac{d}{dt}\ave{\pv} = -m\omega_0^2\ave{\rv}\,.
\end{equation}
Let us now consider the effect of a collision of two test
particles. Of course, the trajectories of the colliding test particles
will not obey any more the original equations of motion
(\ref{eq:eqmotionharmonic}), but the collision has absolutely no
effect on the averages: Since the positions do not change during the
collision, $\ave{\rv}$ remains unchanged, and since the total momentum
of the two colliding test particles is conserved, the average
$\ave{\pv}$ is not changed either. So, the equations of motion
(\ref{eq:eqmotionsloshing}) for the averages $\ave{\rv}$ and
$\ave{\pv}$ remain valid in the presence of collisions. Their solution
is of course an undamped oscillation of the center of mass $\ave{\rv}$
with frequency $\omega_0$. This is confirmed by the numerical result
shown in the upper panel of \Fig{fig:sloshingbreathing}.

\subsection{Breathing mode}
\label{sec:breathing}
A couple of experiments studied the damping of the longitudinal and
radial breathing modes
\cite{KinastHemmer,KinastTurlapov,Bartenstein,Altmeyer_breathing,RiedlBruun}
in elongated traps. In a spherical trap, there is only one breathing
mode (monopole mode), corresponding to an oscillation of the
mean-square radius $\ave{r^2}$ around its equilibrium value
$\ave{r^2}_\eq$. In a spherical harmonic trap, 
this mode is undamped and its frequency $2\omega_0$ is
independent of the number of collisions, like in the case of the
sloshing mode.

Again, this is easy to see. Consider the average kinetic and potential
energies, $\ave{E_\kin} = \ave{p^2}/2m$ and $\ave{E_\pot} =
m\omega_0^2\ave{r^2}/2$. In equilibrium, both are equal (virial
theorem). Now let us assume that the system is compressed or expanded,
such that $\ave{E_\kin} \neq \ave{E_\pot}$. Using again the equations
of motion (\ref{eq:eqmotionharmonic}), one obtains
\begin{gather}
\frac{d}{dt}(\ave{E_\kin}-\ave{E_\pot}) = -2\omega_0^2\ave{\rv\cdot\pv}\,,\\
\frac{d}{dt}\ave{\rv\cdot\pv} = 2(\ave{E_\kin}-\ave{E_\pot})\,.
\end{gather}
Obviously, these two equations describe an undamped oscillation with
frequency $2\omega_0$. Let us now look if they stay valid in the
presence of collisions. Since the collisions do not change the
positions of the particles and conserve the total kinetic energy, it
is clear that $\ave{E_\kin}$ and $\ave{E_\pot}$ are not affected. Now
let us write the difference of $\ave{\rv\cdot\pv}$ before and after a
collision of two test particles $i$ and $j$:
\begin{equation}
\ave{\rv\cdot\pv}^\prime - \ave{\rv\cdot\pv} = \frac{1}{\Ntil}
  \rv_{ij}\cdot(\qv_{ij}^\prime-\qv_{ij})\,,
\end{equation}
where $\qv_{ij}$ and $\qv_{ij}^\prime$ are the relative momenta (e.g.,
$\qv_{ij} = (\pv_i-\pv_j)/2$) before and after the collision. In the
original collision term as written in \Eq{eq:collisionterm}, particles
have to be at the same position to collide, i.e., $\rv_{ij} = 0$, such
that $\ave{\rv\cdot\pv}$ is not changed. In our simulation this is
somewhat different, since the test particles can collide at a distance
of up to $\sqrt{\tilde{\sigma}/\pi}$. This adds a small noise to
$\ave{\rv\cdot\pv}$. In all practical cases, however, this noise is
completely negligible. As an example we show in the lower panel of
\Fig{fig:sloshingbreathing} the oscillation of the mean-square radius
of the cloud as a function of time. As one can see, it is a perfectly
undamped harmonic oscillation with frequency $2\omega_0$.

\subsection{Excitation of an arbitrary mode}
\label{sec:excitation}
For the theoretical investigation of collective modes, it is
convenient to consider a system which is in equilibrium until it is
excited by a short pulse at $t = 0$. Formally, this is achieved by
adding to the time-independent trap potential a perturbation
term of the form
\begin{equation}
\label{eq:deltapulse}
V_1(\rv,t) = \hat{V}_1(\rv) \delta(t) \,.
\end{equation}
The reason for this choice, which is of course different from the
experimental way of exciting a collective mode, is the following:
Provided the perturbation $\hat{V}_1$ is small enough (such that
the system reacts linearly to it), the response to a perturbation with
arbitrary time dependence, $V_1(\rv,t) = \hat{V}_1(\rv)F(t)$, can easily
be obtained by folding the result for the perturbation
(\ref{eq:deltapulse}) with the function $F(t)$.

By integrating the Boltzmann equation over the (infinitesimal)
duration of the pulse, one can show that the effect of the
perturbation (\ref{eq:deltapulse}) is to change the distribution
function as
\begin{equation}
f(\rv,\pv,0^+) = f(\rv,\pv+\nablav \hat{V}_1(\rv),0^-)\,,
\end{equation}
where $0^+$ and $0^-$ denote the limits $t\to 0$ from above and below,
respectively. In the numerical simulation, this means that all test
particles get a kick at $t = 0$,
\begin{equation}
\pv_i(0^+)=\pv_i(0^-)-\nablav \hat{V}_1(\rv_i(0))
\end{equation}
whereas their positions are not changed by the perturbation.

\subsection{Quadrupole mode}
\label{sec:quadrupole}
From now on we will study the quadrupole mode as an example for a
collective mode with non-trivial properties. We write the perturbation
as
\begin{equation}
\hat{V}_1(\rv) = \frac{c}{2} (x^2-y^2)\,,
\label{eq:potexcquad}
\end{equation}
corresponding to a kick at $t = 0$ of $p_x(0^+)=p_x(0^-)-c x(0)$ and
$p_y(0^+)=p_y(0^-)+c y(0)$. The parameter $c$ determines the amplitude
of the perturbation. If $c$ is chosen too small, it is difficult to
separate the oscillation of the mode from fluctuations; if it is
chosen too large, one is not in the linear-response regime. All the
following results were obtained with $c = 0.2m\omega_0$, corresponding
to moderate amplitudes. By varying $c$ within reasonable limits, we
checked that the amplitude of the resulting oscillation scales
linearly with $c$.
\begin{figure}
\includegraphics[width=8cm]{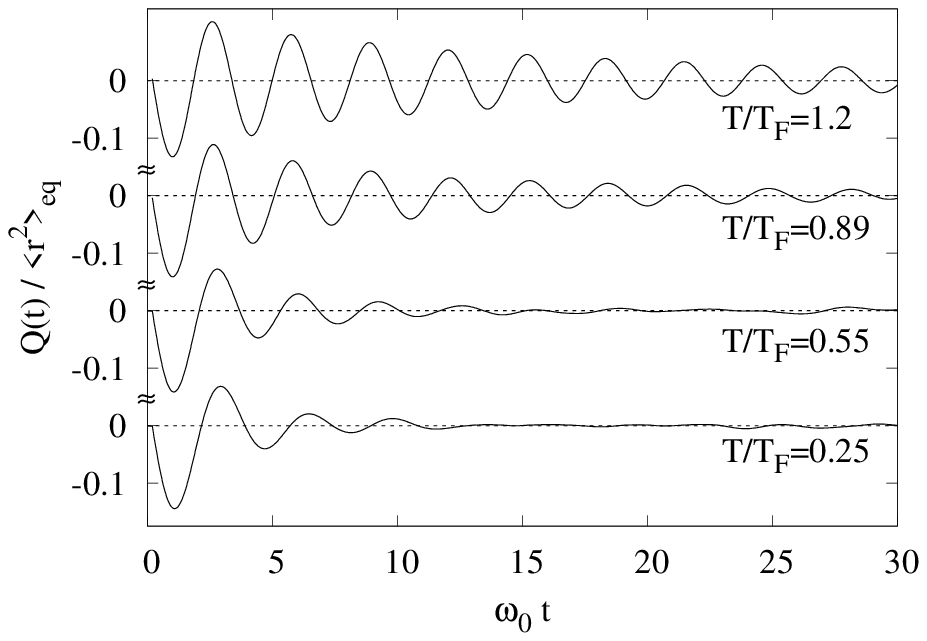}\\
\includegraphics[width=8cm]{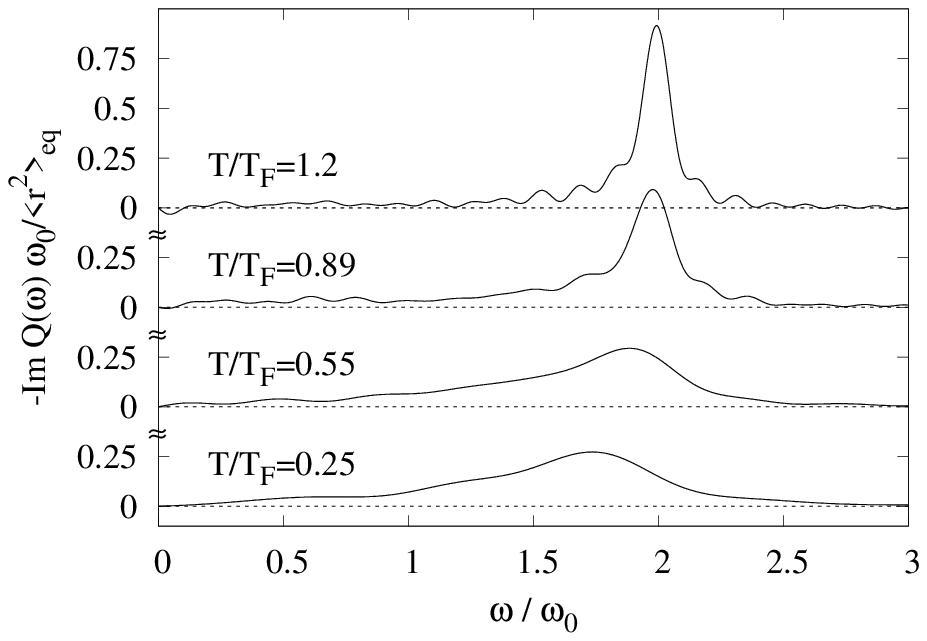}
\caption{Upper panel: Quadrupole response to a perturbation of the
  form (\ref{eq:potexcquad}) with $c = 0.2m\omega_0$ for different
  temperatures. The system has $N=10000$ atoms and
  $1/k_Fa=-0.5$. Lower panel: imaginary part of the corresponding Fourier transforms.}
\label{fig:respmode}  
\end{figure}

After the excitation of the radial quadrupole mode, we can look at the
time evolution of the quadrupole moment $Q = \ave{x^2} - \ave{y^2}$ as
a function of time. Results for different temperatures are displayed
in \Fig{fig:respmode}. Contrary to the sloshing and breathing modes,
the quadrupole mode is damped and the system approaches equilibrium
($Q \to 0$) after a certain time. At high temperatures ($T/T_F \gtrsim
1$), the system gets so dilute that it is in the collisionless regime
($\omega_0 \tau_\coll \gg 1$, $\tau_\coll$ being the mean time between
collisions of one atom). In this case, it takes many oscillations
before the system returns to equilibrium. For lower temperatures, the
mode is damped because of the high collision rate ($\omega_0\tau_\coll
\sim 1$), but the system is not yet in the hydrodynamic regime
($\omega_0 \tau_\coll \ll 1$) where the mode would become undamped
again.

For the analysis of the results, it is useful to apply a Fourier
transform
\begin{equation}
Q(\omega) = \int_0^\infty dt Q(t) e^{i\omega t}\,.
\end{equation}
The so-called response function is the imaginary part of $Q(\omega)$
and can easily be obtained from the numerical results for $Q(t)$ by
using a fast Fourier transform (FFT) algorithm
\cite{numericalrecipes}. As an example, the Fourier transforms of the
results discussed above are shown in the lower panel of
\Fig{fig:respmode}. From the Fourier transform one can clearly see
that the spectrum of the mode in the collisionless regime, i.e., at
high temperature, has a sharp maximum at $\omega = 2\omega_0$, as it
should be in an ideal Fermi gas, whereas at lower
temperature the spectrum is broadened and the centroid of the spectrum
is shifted to lower frequencies. This can be understood since at lower
temperature the system is closer to the hydrodynamic regime, where the
frequency should be $\omega = \sqrt{2}\omega_0$.

Of course, one would like to give numbers $\omega_q$ and $\Gamma_q$
corresponding to the frequency and damping rate of the quadrupole mode
in order to quantify these effects. The simplest way to obtain such
numbers would be to fit the response function $Q(t)$ with a damped
oscillation of the form $-A e^{-\Gamma_q t}\sin\omega_q t$. However,
in the case of strong damping, this ansatz fits very badly the numerical
results for $Q(t)$. This can be understood by looking at the Fourier
transforms: The Fourier transform of this ansatz function is a
Lorentzian, which has a line shape quite different from that obtained
in our numerical simulation for $T/T_F = 0.25$ or $0.55$, cf. lower
panel of \Fig{fig:respmode}. Hence, in order to analyze our numerical
results, we need some physically motivated ansatz for the fit.

\subsection{Comparison with the method of moments}
\label{sec:theodesc}
In most of the theoretical work on collective modes in normal-fluid
Fermi gases, the Boltzmann equation was not solved numerically, but
approximate analytical solutions were found with the help of the
method of moments
\cite{MassignanBruun,BruunSmith,RiedlBruun,Chiacchiera}. For a
detailed description of the method, see e.g. \Ref{Chiacchiera}.

Applying the method of moments to the case of a perturbation of the
form (\ref{eq:deltapulse}) with $\hat{V_1}$ according to
\Eq{eq:potexcquad}, one obtains a theoretical prediction for the
response function $\Im Q(\omega)$. A brief description of the derivation
is given in Appendix \ref{app:mmresp}, the final result reads
\begin{equation}
\Im Q(\omega) = -c \frac{8\ave{E}}{3m^2} \frac{\omega
  \tau}{(\omega^2-2\omega_0^2)^ 2+\omega^2\tau^2(\omega^2-4\omega_0^2)^2}\,,
\label{eq:imQmoment}
\end{equation}
where $\ave{E} = m\omega_0^2\ave{r^2}$ is the mean energy per atom in
equilibrium, and $\tau$ is the relaxation time as defined in
\Refs{BruunSmith,Chiacchiera}, and depends on the cross section (i.e.,
the interaction strength), and the equilibrium distributions,
cf. \Eq{eq:deftau}. One can see from \Eq{eq:imQmoment} that in the
collisionless and hydrodynamic limits the quadrupole mode has the
frequencies $\omega=2\omega_0$ and $\omega=\sqrt{2}\omega_0$,
respectively. The shape of the response function is completely
determined by a single parameter, $\tau$.

By looking for the poles of \Eq{eq:imQmoment}, one can calculate the
inverse Fourier transform which gives $Q(t)$. The result has the form
\begin{equation}
Q(t) = -A e^{-\Gamma_q t}\sin\omega_q t
  +B (e^{-\Gamma_q t}\cos\omega_q t-e^{-\Gamma_1 t})\,,
\label{eq:Qmoment_t}
\end{equation}
i.e., it is a superposition of a damped oscillation with frequency
$\omega_q$ and damping $\Gamma_q$, and a non-oscillating,
exponentially decaying term. The explicit expressions for $\Gamma_1$,
$\Gamma_q$, and $\omega_q$ as functions of $\tau$ as well as for the
amplitudes $A$ and $B$ are given in Appendix \ref{app:qmom}. We will
refer to $\omega_q$ and $\Gamma_q$ as the frequency and damping rate
of the quadrupole mode. Note that in experiments determining these
quantities, the data are usually fitted with a function that is similar
to \Eq{eq:Qmoment_t} \cite{Altmeyer_quadrupole}.

\begin{figure}
\includegraphics[width=8cm]{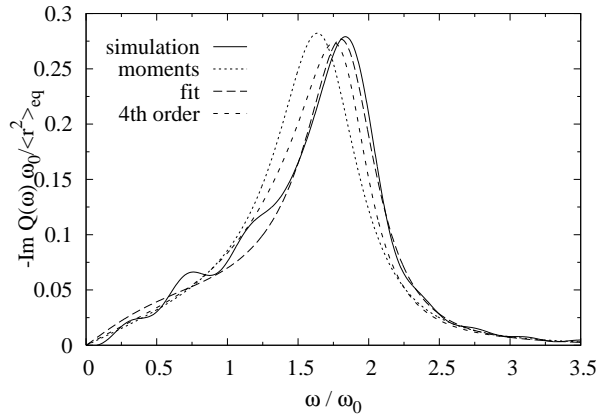}
\caption{Fourier transform of the numerical simulation of the
  quadrupole response (solid line) compared with \Eq{eq:imQmoment}
  where the parameter $\tau$ is obtained by the method of moments
  (dots), or by a fit to the simulation (long dashes). The short
  dashes represent the response obtained with the extended method of moments
  including fourth-order moments. The system is a gas of $N=10000$
  particles at $T=0.4 \, T_F$ and $1/k_Fa = -0.5$.}
\label{fig:fitTF}  
\end{figure}

In \Fig{fig:fitTF} we compare the response function obtained from the
numerical simulation (solid line) with the result obtained from the
method of moments, \Eq{eq:imQmoment} (dotted line). As one can see,
the height of the peak and its general shape are in good agreement,
but the position of the maximum is at different frequencies. However,
if we try to fit the numerical result with a function of the form of
\Eq{eq:imQmoment}, using $\tau$ as fitting parameter, we can very well
reproduce the numerical response function (long-dashed line). It is
remarkable that by adjusting only one parameter, $\tau$, one can
simultaneously reproduce the position, the height and the width of the
peak, and also the shape far away from the maximum. However,
surprisingly, the fitted value of $\tau$ is larger by approximately
$30\%$ than the one obtained by the method of moments. As a consequence,
the frequency $\omega_q$ and damping rate $\Gamma_q$ obtained from the
fit of the response function deviate significantly from those
predicted by the method of moments. These results are summarized in
Table \ref{tab:omegamma}.
\begin{table}
\caption{\label{tab:omegamma} Relaxation time, frequency, and damping
  of the quadrupole mode as obtained from the method of moments and
  fitting the results of the numerical simulation with a function
  of the form (\ref{eq:imQmoment}), corresponding to the dotted and
  dashed curves in \Fig{fig:fitTF}.}
\begin{ruledtabular}
\begin{tabular}{cccc}
method & $\omega_0 \tau$ & $\omega_q/\omega_0$ & $\Gamma_q/\omega_0$ \\
\hline
moments       & 0.451  &   1.676    &   0.353    \\
simulation    & 0.587  &   1.787    &   0.336    \\
\end{tabular}
\end{ruledtabular}
\end{table}

In the existing literature
\cite{MassignanBruun,BruunSmith,RiedlBruun,Chiacchiera}, the method of
moments was limited to second-order moments, as described in appendix
\ref{app:mmresp}. However, as we have seen above, this implies that
the system is characterized by a single relaxation time $\tau$,
whereas in the spirit of a local-density approximation one would
expect that in a trapped system the relaxation time should be position-dependent, 
$\tau = \tau(\rv)$. For instance, one could imagine that
the gas in the center of the trap is more or less hydrodynamic (short
relaxation time), whereas far away from the trap center it gets very
dilute and hence collisionless (long relaxation time). In the case of
the quadrupole mode, this means that the Fermi-surface deformation is
stronger at larger radii than in the trap center. It seems therefore
natural to include into the ansatz for the perturbed distribution
function in addition to the standard term $\propto p_x^2-p_y^2$
describing the Fermi-surface deformation, a term $\propto r^2
(p_x^2-p_y^2)$. More generally speaking, we should go
beyond the standard approximation to include only second-order
moments, and include also fourth-order (or perhaps even higher)
moments.

\begin{figure}
\includegraphics[width=8cm]{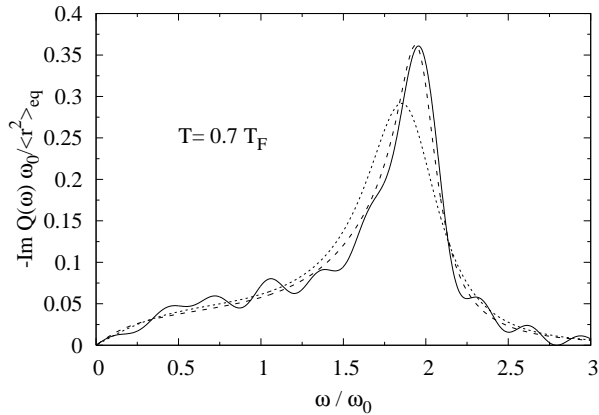}
\caption{Same as \Fig{fig:fitTF} for a temperature $T=0.7 \, T_F$. 
For clarity, the fit of the simulation is not shown. The system is a gas of $N=10000$
  particles at $1/k_Fa = -0.5$.}
\label{fig:compfourth0.7}  
\end{figure}

The task of extending the method of moments to the next higher order
is in principle straight-forward but in practice very tedious: In the
case of the quadrupole mode, the number of moments is increased from
three to twelve. Some details are given in appendix
\ref{app:fourthorder}. The resulting response function is shown in
\Fig{fig:fitTF} as the short-dashed line. Surprisingly, its shape is
still similar, but now the position of the maximum agrees rather well
with the result of the numerical simulation (solid line). The agreement is even
better at higher temperature (see \Fig{fig:compfourth0.7}).
This nicely
confirms the correctness of our numerical simulation and shows
explicitly that the method of moments, if truncated at the lowest
order, is insufficient.

\begin{figure}
\includegraphics[width=7cm]{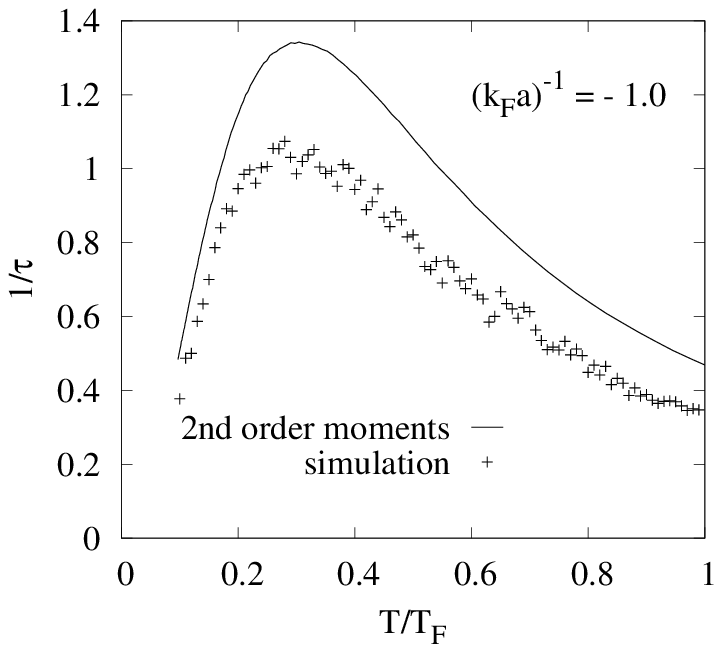}
\includegraphics[width=7cm]{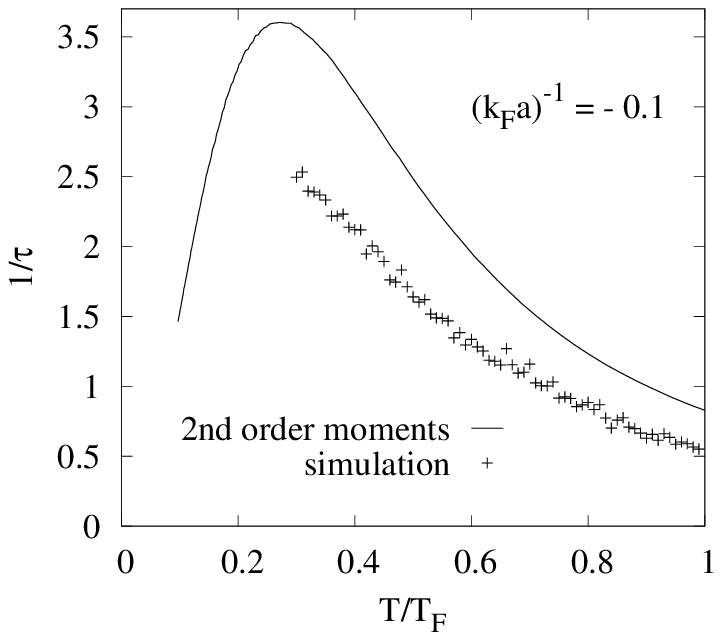}
\caption{Comparison of the inverse relaxation time, $1/\tau$, as obtained from the
simulation (crosses) and from the method of moments \Eq{eq:deftau}
(solid lines), as a function of temperature. The system consists of $N =
10000$ atoms with $1/k_Fa = -1$ (upper panel) and $-0.1$ (lower panel).}
\label{fig:tauinv}  
\end{figure}

By doing
calculations for various interaction strengths and temperatures, we
found that the relaxation time from the simulation is systematically
longer than that from the method of moments (without fourth-order
moments), \Eq{eq:deftau}. Results for weaker and stronger interactions
($1/k_F a = -1$ and $-0.1$) are displayed in \Fig{fig:tauinv}.
We see that the general behavior of $\tau$ as a function of
temperature is the same within the simulation and the method of
moments, but quantitatively there is a discrepancy of the order of
$30\%$ in the whole range of temperatures where our numerical
simulation is very accurate ($T > 0.35 \, T_F$, cf. \Fig{fig:collrate}
showing the temperature dependence of the collision rate). Note that
at lower temperatures, the determination of the Pauli-blocking factors
in the simulation of the collisions is not completely accurate, as
discussed below \Fig{fig:collrate}, such that the collision rate below
$0.35 \, T_F$ is slightly too high. Nevertheless the inverse relaxation
time is too small. From this one can conclude that if we could improve
the Pauli blocking in the simulation, the discrepancy between the
simulation and the method of moments (without fourth-order moments)
would be even worse. The fourth order is thus important for the
determination of the relaxation of the system and particularly 
for the frequency and the damping of collective modes.

\subsection{Frequency and damping of the quadrupole mode}
\label{sec:freqdamp}

\begin{figure}
\includegraphics[width=7cm]{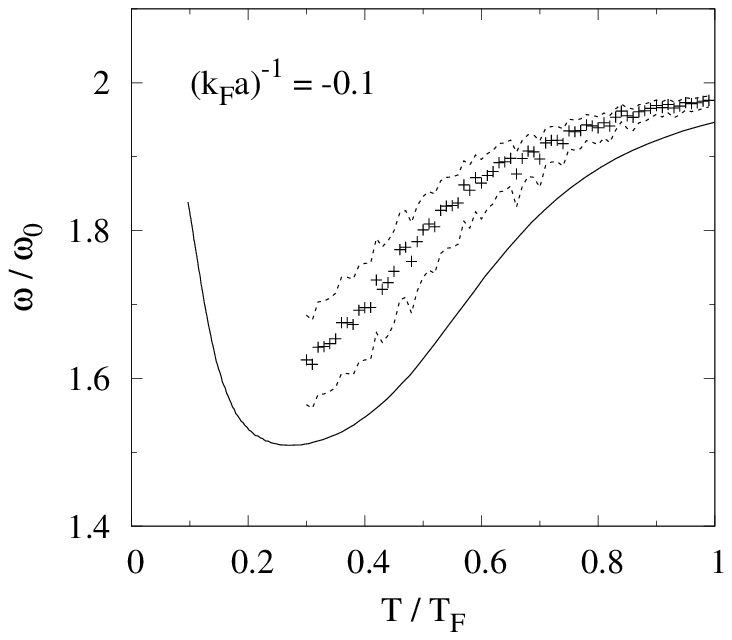}
\includegraphics[width=7cm]{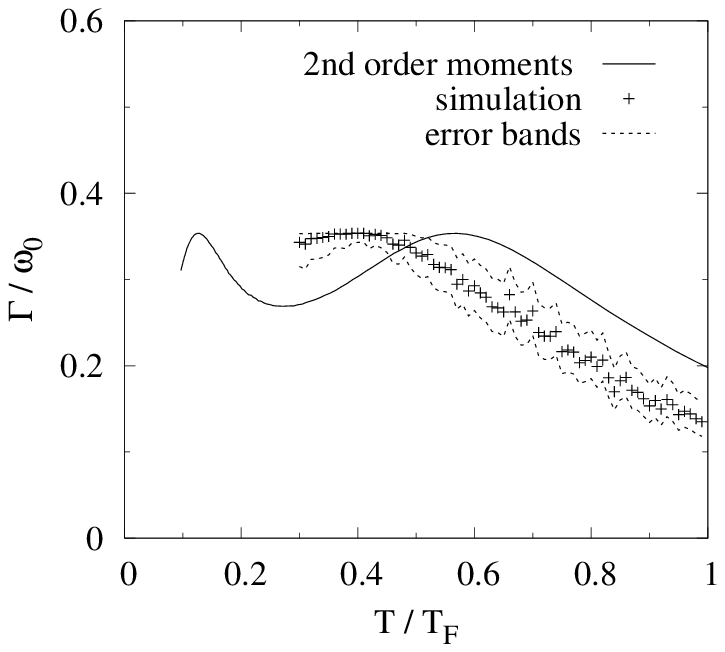}
\caption{Frequency (top) and damping rate (bottom) of the quadrupole
mode as a function of the temperature as obtained from the numerical
simulation (crosses) and from the method of moments (solid line). The
short-dashed lines indicate the error band of the results 
from the simulation if we admit that the relaxation time $\tau$ of the 
simulation may be wrong by 15\%. The system consists of $N=10000$ 
particles close to unitarity ($1/k_Fa=-0.1$).}
\label{fig:quauni}  
\end{figure}

As we have just seen, the numerical simulation gives systematically a
longer relaxation time $\tau$ than the method of moments. As the
frequency $\omega_q$ and damping rate $\Gamma_q$ of the quadrupole
mode are parametrized in terms of $\tau$ (see Appendix \Sec{app:qmom}), 
one can ask the question how
strongly this difference in $\tau$ will affect the results for
$\omega_q$ and $\Gamma_q$. Since we are mainly interested in the
intermediate regime $\omega\tau \sim 1$ between the hydrodynamic and
collisionless limits, a difference of $30\%$ in $\tau$ can completely
change the temperature dependence of $\omega_q$ and $\Gamma_q$. This
is shown in \Fig{fig:quauni}, where the crosses are the results
obtained from the simulation, whereas the solid lines are the results
from the method of moments. One can clearly see that the numerical
results stay close to the collisionless limit to much lower
temperatures than the results obtained by the method of moments.

To estimate the resulting precision of our numerical result on $\omega_q$ 
and $\Gamma_q$, we show  in \Fig{fig:quauni} 
the error bands (short-dashed lines) which we obtain if we assume that our
simulation may give a $\tau$ which is wrong by at most $15 \%$. This error
includes numerical uncertainties which can be estimated from the 
scattering of the points in \Fig{fig:tauinv} and the systematic deviation 
of the collision rate shown in \Fig{fig:collrate}. 

If we include the fourth order moments, a global relaxation time $\tau$ 
does not exist anymore but we could define an effective one by fitting 
the response function as we did for the simulation. This effective relaxation 
time agrees very well with the one of
the simulation so that both results give very similar frequency 
and damping. However, from a theoretical point of view, the
definition of these quantities should come from the zeroes of the 
determinant of the matrix $A_{ij}$ defined in Appendix \ref{app:fourthorder}. Such a 
discussion is postponed to a forthcoming publication \cite{lepers.thesis}.

%
\section{Conclusions}
\label{sec:Conclusions}
In this paper, we presented a test-particle method for solving
numerically the Boltzmann equation for trapped Fermi gases. While such
methods have been popular in other fields of physics for many years,
there have been only a few applications to ultracold atomic gases
\cite{Urban,ToschiVignolo,ToschiCapuzzi,JacksonZaremba}. Our method is
similar to that of \Refs{ToschiVignolo,ToschiCapuzzi} with some differences in the
treatment of the collision term. In order to compute the occupation
numbers in the Pauli-blocking factors in the collision term, we
represent each test particle by a Gaussian in $\rv$ and $\pv$
space. The minimum value of the width of the Gaussian is dictated by
the statistical fluctuations due to the finite number of test
particles, limiting the applicability of the method to temperatures
above $\sim 0.2 \, T_F$.

As a first application of the method we discussed some collective
modes. For simplicity, we considered only toy systems consisting of
$\sim 10^4$ atoms in a spherical harmonic trap and neglected the
mean-field potential and medium modifications of the cross section. As
expected, the sloshing and monopole modes are undamped and independent
of the collisions. In contrast, the quadrupole mode is very sensitive
to collisions. In the hydrodynamic limit, its frequency should
approach $\sqrt{2}\omega_0$, while it is $2\omega_0$ in the
collisionless limit. In our simulations, we never reach the
hydrodynamic regime, but the collisionless regime can be realized at
high temperature due to the diluteness of the gas.

Surprisingly, the frequency and damping rate of the quadrupole mode
obtained within the numerical simulation are quite different from
those obtained within the widely used method of moments including
moments up to second order in $\rv$ and $\pv$. The method of moments
predicts a relaxation time $\tau$ which is significantly shorter than
the one obtained within the simulation. The reason is that the $\rv$
dependence of the relaxation time is neglected if only the
$p_x^2-p_y^2$ moment is taken into account for the description of the
Fermi-surface deformation. We have shown that if the method of moments
is extended to moments up to fourth order in $\rv$ and $\pv$, e.g.,
the $r^2(p_x^2-p_y^2)$ moment, the agreement with the simulation
becomes very good.

The focus of the present paper was mainly to explain the test-particle
method and to show its usefulness. For instance, the deficiency of the
method of moments up to second order would not have been detected
without the comparison with the numerical result. In future studies,
we plan to apply the method to more realistic cases. In particular, in
order to reach the typical numbers of atoms in the experiments, we
will have to increase $N$ by a factor of $\sim 10-100$. However, this
should not pose a big problem: According to \Eq{conditionT}, if we
increase $N$ but keep the ratio $T/T_F$ fixed, the widths $w_r$ and
$w_p$ may be chosen larger ($\propto N^{1/6}$). This means that
\Eq{conditionN} stays satisfied with the same number of test
particles, $\Ntil$, i.e., with a reduced ratio $\Ntil/N$. The
computation time will only grow because of the increased collision
rate (due to the larger test-particle cross section $\tilde{\sigma} =
\sigma N/2\Ntil$). Another point is the trap geometry. The traps in
the experiments are usually not spherical, but elongated. Concerning
the propagation of the test particles, this does not cause any
difficulty, but in the calculation of the occupation numbers, it will
probably be necessary to replace the width $w_r$ of the Gaussian in
$\rv$ space by different widths $w_x$, $w_y$, and $w_z$ in the three
space directions. Another important advantage of the numerical method is
that an anharmonicity of the trap potential, which is always present
in real experiments, can easily be included.

Finally, the mean field \cite{Chiacchiera} and medium modifications of
the cross section \cite{RiedlBruun,Chiacchiera} should be
included. The mean field, which originally depends on the chemical
potential $\mu$ and the temperature $T$, can be expressed as a
function of the local density and energy density, which are both
obtainable in the simulation. However, as shown previously 
\cite{Chiacchiera}, 
the mean field is not just proportional to the density: this leads 
to a huge numerical effort which is beyond the scope of this paper.
The in-medium cross section is also 
difficult to be included 
because it depends on too many variables to be tabulated:
$\sigma = \sigma(k=|\pv+\pv_1|/2,q=|\pv-\pv_1|;\mu,T)$. One possible
solution of this problem is to replace the full $k$ and $q$ dependence
of the in-medium cross section by a simple parametrization which
results in the same local relaxation time $\tau(\mu,T)$. Work in this
direction is already in progress. In \Refs{RiedlBruun,Chiacchiera},
within the method of moments up to second order, the use of the
in-medium cross section spoiled the agreement with experimental data
because the resulting relaxation times were too short. Since the
present work shows that the numerical simulation gives a longer
relaxation time than the method of moments, we hope that this problem
can be solved.

Further important extensions of the present work are the
generalization to polarized Fermi gases and to superfluid
systems. These questions, however, require more fundamental
theoretical studies before they can be tackled numerically.
\appendix
\section{Collision rate at equilibrium}
\label{app:collrate}
Replacing the
distribution functions in \Eq{eq:collrate} by equilibrium distribution
functions $f_{\eq}$, one obtains after some algebra the equilibrium collision
rate
\begin{multline}
\dot{N}_{\coll,\eq} = \frac{1}{4\pi^4} \int d^3 r\,\int_0^\infty
  dk\,k^2 \int_0^\infty dq\,q^2 \frac{2q}{m} \sigma(q)\\ \times\left(
  \frac{\tanh^{-1}(\tanh\frac{X}{2}\tanh\frac{Y}{2})} {Y \sinh X}
  \right)^2\,,
\label{eq:MCblock}
\end{multline}
where $\kv=\pv+\pv_1$, $\qv=(\pv-\pv_1)/2$,
$X=\beta(k^2/8m+q^2/2m+V_T-\mu)$ and $Y=\beta kq/2m$.
The total rate of allowed and blocked
collisions is, in turn, given by
\Eq{eq:collrate} but without the factor
$(1-f^{\prime})(1-f_1^{\prime})$ in the integrand, leading to
\begin{multline}
\dot{N}_{\coll,\eq}^\mathit{(+blocked)} = \frac{1}{4\pi^4}
  \int d^3 r\int_0^\infty dk\,k^2\int_0^\infty dq\,q^2\frac{2q}{m}\sigma(q)\\ 
  \times \frac{\tanh^{-1}(\tanh
  \frac{X}{2}\tanh \frac{Y}{2})}{Y e^X\sinh X}\,.
\label{eq:MCnoblock}
\end{multline}
If the trap potential is spherically symmetric or harmonic, the
spatial integrals can be reduced to one-dimensional ones. The
remaining three-dimensional integrals are evaluated numerically with a
Monte-Carlo algorithm.
\section{Quadrupole response within the method of moments}
\label{app:mmresp}
In the case of a weak perturbation, we can write the deviation of the
distribution function from the equilibrium one in the form
\begin{equation}
f-f_\eq = 
  f_\eq(1-f_\eq)\Phi(\rv,\pv,t)\,.
\label{eq:f1Phi}
\end{equation}
Inserting this expression into the Boltzmann equation and keeping only
terms linear in the perturbation, one obtains (see Eq.\@ (36) of
\Ref{Chiacchiera}, for the case without mean field but with an
external perturbation):
\begin{multline}
f_\eq(1-f_\eq) \Big(\dot{\Phi} +
  \frac{\pv}{m}\cdot\nablav_r\Phi -
  \nablav_r V_T \cdot\nablav_p\Phi \\
  + \beta\frac{\pv}{m}\cdot\nablav_r V_1
\Big) = -I[\Phi].
\label{eq:BoltzLin}
\end{multline}
Here, $I[\Phi]$ is the linearized collision term as defined in Eq.\@
(37) of \Ref{Chiacchiera} (up to a factor $(2\pi)^3$ since here we are
using a different normalization of $f$):  
\begin{multline}
I[\Phi] = \int \frac{d^3p_1}{(2 \pi )^3}\int d\Omega
\frac{d\sigma}{d\Omega} |\vek{v}-\vek{v_1}|
  f_\eq f_{\eq\,1}\\
  \times (1-f_\eq^\prime)(1-f_{\eq\,1}^\prime)
  (\Phi+\Phi_1-\Phi^\prime-\Phi_1^\prime)\,.
\label{collisionterm}
\end{multline}    
The perturbation $V_1$ is given by
Eqs. (\ref{eq:deltapulse}) and (\ref{eq:potexcquad}). 
The usual approximation consists in making the ansatz
\begin{equation}
\Phi(\rv,\pv,t) = \sum_{i=1}^3 c_i(t) \phi_i(\rv,\pv)\,,
\label{eq:ansatzmom2}
\end{equation}
with time-dependent coefficients $c_i$ and $\phi_1 = x^2-y^2$, $\phi_2
= xp_x-yp_y$, and $\phi_3 = p_x^2-p_y^2$, i.e., only quadratic moments
are considered. Evaluating the moments $\int d^3r d^3p
\phi_i(\rv,\pv)\times \mbox{\Eq{eq:BoltzLin}}$, one obtains a system of
equations for the Fourier transformed coefficients:
\begin{equation}
\sum_{j=1}^{3} A_{ij} c_j(\omega) = a_i
\end{equation}
with
\begin{multline}
A_{ij} = \int\frac{d^3r d^3p}{(2\pi)^3} \phi_i \Bigg\lbrack f_\eq(1-f_\eq) \\
\times \left( -i\omega \phi_j + \left\lbrace \phi_j, \frac{p^2}{2m}+\frac{m\omega_0^2r^2}{2} \right\rbrace \right) 
+ I[\phi_j] \Bigg\rbrack
\end{multline} 
and
\begin{equation}
a_i = - \beta \int\frac{d^3r d^3p}{(2\pi)^3} \phi_i f_\eq(1-f_\eq) \frac{\pv}{m} \cdot \nablav \hat{V}_1 \,,
\end{equation}
where $\lbrace .,. \rbrace$ are the Poisson brackets. Using the virial
theorem, we obtain explicitly:
\begin{gather}
-i \omega c_1 -  m\omega_0^2 c_2 = 0 \\
\label{eq:system}
2 c_1-im \omega c_2 - 2m^2\omega_0^2 c_3 = - \beta c\,,\\
c_2+\Big(\frac{1}{\tau}-i \omega\Big)mc_3 = 0 \,,
\end{gather}
where the relaxation time $\tau$ is defined by
\cite{RiedlBruun,Chiacchiera}
\begin{equation}
\frac{1}{\tau} = \frac{3\beta}{m^2 N\ave{E_\kin}} 
  \int\frac{d^3r d^3p}{(2\pi)^3} p_x p_y I[p_x p_y]\,.
\label{eq:deftau}
\end{equation}
Solving this system of equations, we find
\begin{equation}
c_1(\omega) = \frac{\beta\omega_0^2 c (1-i \omega \tau)}
  {\omega^2-2\omega_0^2-i\omega \tau(\omega^2-4\omega_0^2)}\,.
\label{eq:c1omega}
\end{equation}
and similar expressions for $c_2$ and $c_3$. However, only the
coefficient $c_1$ contributes to $Q$: With \Eq{eq:f1Phi}, and using
again the virial theorem, we obtain $Q(t) = \ave{x^2-y^2} =
4T\ave{r^2}c_1(t)/3m\omega_0^2$, or explicitly:
\begin{equation}
Q(\omega) = \frac{4\ave{r^2}c}{3m}\, \frac{1-i\omega\tau}
     {\omega^2-2\omega_0^2-i\omega \tau(\omega^2-4\omega_0^2)}\,.
\label{eq:Qomega}
\end{equation}
Taking the imaginary part, we obtain \Eq{eq:imQmoment}.
\section{Time dependence of the quadrupole response within the method of
  moments}
\label{app:qmom}
In order to compute the Fourier transform of \Eq{eq:Qomega}, let us
start by factorizing the denominator:
\begin{multline}
\omega^2-2\omega_0^2-i\omega \tau (\omega^2-4\omega_0^2)\\
  = -i\tau(\omega-\omega_1)(\omega-\omega_2)(\omega-\omega_3)\,.
\end{multline}
The expressions for the roots $\omega_i$ can be given in closed
form. Defining $\tautilde = \omega_0\tau$ and
\begin{gather}
\Theta = \left( 1+9 \tautilde^2+3\tautilde
\sqrt{6-39\tautilde^2+192\tautilde^4}\right)^{1/3}\,,\\
u_\pm = \frac{1}{3\tau}\left(\Theta \pm 
  \frac{1-12\tautilde^2}{\Theta}\right)\,,
\end{gather}
we can write the roots $\omega_i$ as
\begin{equation}
\omega_1 = -i\Gamma_1\,,\quad
\omega_2 = \omega_q-i\Gamma_q\,,\quad
\omega_3 = -\omega_q-i\Gamma_q\,,
\end{equation}
with
\begin{equation}
\Gamma_1 = \frac{1}{3\tau}+u_+\,,\quad
\Gamma_q = \frac{1}{3\tau}-\frac{u_+}{2}\,,\quad
\omega_q = \frac{\sqrt{3}}{2}u_-\,.
\end{equation}
Now it is straight-forward to evaluate the inverse Fourier transform
of the response (\ref{eq:imQmoment}) using the residue theorem. The
result is given by \Eq{eq:Qmoment_t} with
\begin{gather}
A = \frac{4c\ave{r^2}}{3m\omega_q\tau}\, \frac{\omega_q^2\tau+
  (\Gamma_1-\Gamma_q)(1-\Gamma_q\tau)}
  {\omega_q^2+(\Gamma_q-\Gamma_1)^2}\,,\\
B = \frac{4c\ave{r^2}}{3m\tau}\, \frac{1-\Gamma_1\tau}
  {\omega_q^2+(\Gamma_q-\Gamma_1)^2}\,.
\end{gather}
\section{Extension of the method of moments to fourth-order moments}
\label{app:fourthorder}

Taking fourth order moments into account, we extend the previous ansatz 
\Eq{eq:ansatzmom2} as follows:
\begin{eqnarray}
\Phi & = & c_1 (x^2-y^2) +c_2 (xp_x-yp_y)+c_3 (p_x^2-p_y^2) \nonumber \\
& + & c_4 r^2(x^2-y^2)+c_5 p^2(x^2-y^2)+c_6 \rv \cdot \pv (x^2-y^2) \nonumber \\
& + & c_7 r^2(xp_x-yp_y) +c_8 p^2(xp_x-yp_y) \nonumber \\
& + & c_9 \rv \cdot \pv (xp_x-yp_y) + c_{10} r^2 (p_x^2-p_y^2) \nonumber \\
& + & c_{11} p^2 (p_x^2-p_y^2)+ c_{12} \rv \cdot \pv(p_x^2-p_y^2)
\end{eqnarray}
which can be written as $\Phi = \sum_{i=1}^{12} c_i \phi_i $
with, for example, $\phi_1 = (x^2-y^2)$.
Following the same steps as explained in Appendix \ref{app:mmresp}, we obtain 
now a system of twelve equations. The matrix $A_{ij}$ can be computed explicitly. 
Contrary to the second order calculations, the virial theorem can no longer be used to
reduce the number of unknown quantities so that the system now depends on
$\langle r^2 \rangle$, $\langle r^4 \rangle$ and $\langle r^6 \rangle$.
In the matrix elements of the collision term, more parameters appear, generalizing
the single parameter $\tau$ of the second order method.

After solving the system of equations numerically, we can express the quadrupole
moment in terms of the coefficients $c_i$ as :

\begin{equation}
Q(\omega) = \frac{4T}{3} \left\lbrack \frac{\langle r^2 \rangle}{m \omega_0^2} c_1
+ \frac{\langle r^4 \rangle}{5}\left( \frac{7 c_4}{m \omega_0^2} + 3 m c_5 + m c_9 \right) \right\rbrack \,.
\label{eq:qmomsup}
\end{equation}

Further details and explicit formula for the matrix can be 
found in \cite{lepers.thesis}.


\end{document}